# GROWTH-ALGORITHM MODEL OF LEAF SHAPE

David A. Young
Lawrence Livermore National Laboratory
Mail Stop L-45
7000 East Avenue
Livermore, California 94550
email address: young5@llnl.gov
April 2010

## ABSTRACT

The innumerable shapes of plant leaves present a challenge to the explanatory power of biophysical theory. A model is needed that can produce these shapes with a small set of parameters. This paper presents a simple model of leaf shape based on a growth algorithm, which governs the growth rate of leaf tissue in two dimensions and hence the outline of the leaf. The growth of leaf lobes is governed by the position of leaf veins. This model gives an approximation to a wide variety of higher plant leaf shapes. The variation of leaf shapes found in closely related plants is discussed in terms of variability in the growth algorithms. The model can be extended to more complex leaf types.

# I. INTRODUCTION

Modern physics has recently incorporated the problem of complex natural phenomena as a strong research focus. This has brought problems of biological process and structure into the domain of theoretical physics. Biological pattern formation or morphogenesis is an important aspect of complexity theory. One example of pattern formation is the origin of leaf shapes. Human admiration of the marvelous variety of leaf shapes is ancient, and has stimulated modern scientific analysis of how leaves attain their shapes. Experimental investigations on the causation of leaf shapes began more than a century ago, and continue today [1]. Genetic components of leaf growth have been discovered and are beginning to provide a picture of how gene products influence leaf shape [2]. However, it is a legitimate question how far molecular-genetic investigations can go in analyzing morphogenetic processes, which inevitably involve multi-cellular organization. Like other morphogenetic processes, leaf growth and shape are poorly understood.

Theoretical studies are valuable in providing models of pattern formation that stimulate new research and that can be tested against experimental data [3,4]. Theory offers the promise of bridging the very large gap between the expression of genes and the final shape of an organ. There are a number of divergent hypotheses explaining leaf form. One of the earliest is the work of Thompson in his famous book, *On Growth and Form*. In a brief section on leaf morphogenesis, Thompson suggests that a polar coordinate function fit to a leaf outline indicates a vector diagram of the growth process [5]. He gives an example of a function that resembles the horse chestnut (genus *Aesculus*) leaf with its many lobes. Other models include the Lindenmayer L-system [6], fractal analysis [7], a Turing reaction-diffusion process [8], an iterative space-filling branching process [9], and a linear force-relaxation model [10]. The formation of the vein network of leaves, which is closely related to leaf growth, has also been studied in recent theoretical work [11,12]. Much more experimental and theoretical work is needed to develop a convincing theory of leaf shape.

In this paper I introduce a very simple model of leaf growth that produces a spectrum of shapes approximating those observed in nature. My objective is to explain not only the range of leaf shapes found in the higher plants, but also the large variations in shape seen in leaves on closely related plants. The results of the model simulations naturally lead to speculations about the combined influence of genes and environment on leaf morphogenesis.

## II. LEAF SHAPES IN NATURE

Leaves have evolved over millions of years to optimize light collection, transport of nutrients to and from the plant body, and mechanical stability against natural stresses. Even under the optimizing force of natural selection, however, leaves are found in innumerable forms, showing that leaf shape is a response to multiple competing influences. Variability of leaf shape within a single genus or even on a single plant is especially interesting, because it indicates that the controlling shape-generating "algorithm" has multiple components that can be independently varied. The study of complex physical structures such as diffusion-limited aggregates, dendritic crystals, etc., has led to the conclusion that they are the result of endlessly iterated simple processes [13]. It is plausible that biological structures can be explained similarly, and that only a few model parameters are needed to generate the variety of structures being considered. The task of the theorist is to decipher the structure-generating algorithm from the various shapes that appear in nature.

Botanists have developed numerous terms for the shapes of leaves as an aid to identifying plant species [14], but these names are not based on a biophysical understanding of leaf growth. It is perhaps now time for a more biologically motivated scheme to explain the variety of shapes.

The model described here was motivated by a small number of well-established facts. First, leaf shape is not rigidly programmed by the genetic material. This becomes obvious when we consider the phenomenon of *heteroblasty*, in which the shapes of the leaves on a single plant change with their time of appearance in the growing shoot [15]. There is clearly a post-genetic or "environmental" influence on leaf shape, which may include such variables as nutrition, light levels, $CO_2$ levels, and growth hormone concentrations. Second, leaf growth is exponential over about three orders of magnitude in linear dimension [16]. This growth pattern implies that the leaf shape is *allometric* [17], meaning that differences in the growth rate of different leaf features produce a constantly changing shape in the growing leaf and therefore a potentially variable mature shape. Third, the study of chimeric leaves, in which mutant cells proliferate to produce visible clones in the mature leaf, has shown tissue growth patterns that are mainly parallel to secondary veins [18]. This provides a basis for the growth direction chosen by the model.

# III. THE MODEL

Let us assume that a leaf is a two-dimensional sheet of growing tissue. The model simulation begins with two parallel columns each with N tissue zones or "cells", with separation distance equal to 1.0. One column (x = 0) is the midrib or primary vein, with height equal to 1.0 and the other (x = 1.0) is the right-hand leaf margin, which may have a variable height. The cells in the two columns are each given an integer index j, where j = 1 at the petiole or point of attachment to the stem, and j = N at the leaf tip. The model works on only half of the leaf, and the symmetric other half is added at the end of the simulation for graphical display. A growth rate algorithm dependent on the j index of each cell is specified with several parameters. In this model growth occurs in the y direction parallel to the midrib, and also in the direction of the cell rows set by the slopes of rows designated as secondary veins. A cell "row" is defined as a pair of midrib and margin cells with the same j value. It is assumed that each row represents a growing file of cells, but it is unnecessary to actually include these cells in the model simulations. At each growth (time) step, the algorithm is applied to each midrib and margin cell, and the position of that cell is changed accordingly. The algorithm specifies a multiplicative factor on the position, so that growth is exponential in time, with the result that the shape evolves allometrically during growth. The initial cell array is allowed to expand by a factor of about 1000 to reach a mature shape. A typical value of N that gives adequate definition of the leaf shape is 150.

Secondary veins are an important component of the growth algorithm. A model vein consists of a cell row designated as a vein in the input file. Veins set the direction of growth of leaf cells and control the formation of lobes. The number, position, and growth direction or slope of veins are set as an initial condition. Veins may be given different slopes at different j values along the midrib, and non-vein cell rows between such veins are linearly interpolated in their slopes. The slope of any cell row is kept constant during growth.

The model "cells" are of course not intended to mimic real biological leaf cells. The model does not increase the cell number with growth, and the balance between cell proliferation and expansion is not included in the model. The only important model growth variables are the changing positions of midrib and margin cells.

The growth in the y-direction is set at a constant rate for each cell. This means that the y position of each midrib cell is increased by a constant percentage at each growth step. For an arbitrary time interval of 1, the position y of any midrib cell at time t + 1 will be increased by a constant multiplicative factor:

$$y(t+1) = (1+g_y)y(t) \tag{1}$$

A typical model growth rate $g_y$ is 10%, so the y position of any midrib cell is then increased by a factor of 1.1. This increment is added to the y position of the margin cell with the same index j, so that the whole array elongates uniformly in the y direction.

The growth in the x direction is by elongation of the cell rows, each with length L dependent on the index j. If the rows have non-zero slope, then both the x and y values of the margin cells change with row elongation:

$$L(t+1) = (1+g_L)L(t) \tag{2}$$

For a cell row with slope s, the increase in length L is given by

$$\Delta L^2 = \Delta x^2 + \Delta y^2 = \Delta x^2 + s^2\Delta x^2 = \Delta x^2(1+s^2)$$

Hence

$$\Delta x = \frac{\Delta L}{(1+s^2)^{1/2}}$$

$$\Delta y = s\Delta x$$

As shown if Fig. 1, the increase in x position occurs by row elongation. The increase in y position occurs both by uniform axial growth and by row elongation if the row has nonzero slope. The elongation rate for each cell row is given as a function of the position j of the row along the midrib. The cell row elongation growth formula has a maximum value at an index $j_d$. For $j < j_d$ or $j > j_d$, the growth rate drops off, and is zero at the endpoints. Motivated by the common ovate shape of many leaves, I propose a simple power law dependence of the growth rate on j:

$$1 \le j \le j_d \qquad g_L = g_{L0}\left[1-\left(\frac{j_d - j}{j_d - 1}\right)^k\right] \qquad (3a)$$

$$j_d < j \le N \qquad g_L = g_{L0}\left[1-\left(\frac{j - j_d}{N - j_d}\right)^k\right] \qquad (3b)$$

These formulas guarantee that the growth rate is equal to $g_{L0}$ at the boundary point $j_d$ and that it falls off to zero at the endpoints. The exponent k determines how this falloff occurs. It can be slow if k is large or fast if k is small. The parameters $y_d = j_d/N$ and k have a strong effect on the leaf shape.

Eq. (3) represents a *positional information* model [3] in which a spatial pattern is generated that differentially influences cellular behavior. Positional information normally refers to tissue differentiation, but here it refers to growth rate. The growth rate $g_L$ is constant for any index j, that is, the relative position, even though the absolute size of the leaf has grown by a large factor. I assume that the growth rate of each cell row is "imprinted" with its relative position in the embryonic leaf primordium, as indicated by the index j, and that this is transmitted to each cell's many descendants.

For dissected leaves with lobes, I make the assumption that the lobes are centered on the initially established veins. Cell rows designated as veins have the full growth rate $g_L$. Non-vein cell rows more distant from the veins will have a lower growth rate determined by the index distance to the nearest vein, $|j - j_v|$. The algorithm for each cell row finds the nearest vein and computes the index distance. Then the growth rate $g_L$ in Eq. (3) is multiplied by the factor

$$f = \exp(-a\,|\,j - j_v\,|^n) \qquad (4)$$

Two new parameters *a* and *n* are introduced to determine the shape of the leaf lobes. This form is suggested by the observations that to a good approximation leaf lobes are centered on secondary veins and that the shape of the indentations between neighboring veins is symmetrical with respect to the vein positions. Thus in the model the farther the cell row is from the nearest vein, the lower its elongation rate, leading to a dissected shape.

A model simulation begins with specification of the vein positions $j_v$ and their slopes s and the choice of the growth parameters $y_d$, k, $g_y$, $g_{L0}$, *a*, and n. Veins are treated as ordinary cell rows in the growth algorithm, but are centers of growth in dissected leaves. The growth rate is calculated for each midrib and margin cell and those cells are given new positions accordingly. The initial leaf or primordium length and half-width are set at 1.0, and the growth steps are iterated until the length of the leaf is about 1000, based on the observation that a typical leaf increases in size from 0.1 mm to 100 mm [16]. If the $g_y$ and $g_{L0}$ rates are different, the leaf will change its (x, y) aspect ratio with growth. These rates are chosen to produce a realistic shape as determined by observation.

There is obviously a monotonic relation between the growth rates $g_y$ and $g_L$ and the final leaf shape. This is illustrated in Fig. 2 for a typical model leaf. The exponential growth means that the final shape is proportional to an exponential function of the growth rate. This amplifies the leaf width in the region of maximum growth rate and suppresses it in regions of low growth rate. Hence the leaf shape will be very sensitive to the choice of the growth rate formula.

## IV. RESULTS

With trial and error parameter changes, the model can generate leaf shapes approximating those observed in real plants. My primary interest is to explain the normal ranges and variations in leaf shapes by continuous changes in the parameters described above. If this can be done, then different leaf shapes can be referred back to parameters in the growth algorithm, which adds biological insight and is a considerable conceptual simplification.

The simulations begin with entire leaves ($a = 0$) which have a smooth outline and then go to dissected leaves ($a > 0$) with more complex shapes. The common and scientific names of plants with leaf shapes similar to the model shapes are mentioned in the figure captions.

In Fig. 3, the ratio R of growth rate in the y direction to that of the L direction, R = $g_y / g_{L0}$ , is varied. In Figs. 3(a) – 3(d), $g_y$ is fixed and $g_{L0}$ is varied. For small R, the leaf is broad, and as R increases, the leaf narrows. The range of R shown here is 1.07 (Fig. 3(a)) to 1.71 (Fig. 3(d)). Although botanists would give independent names (broadly ovate, ovate, broadly lanceolate, and lanceolate) to these forms, it is important to realize that they can all be generated by the change in the single parameter R, and that there is no essential difference in the morphogenetic process creating them.

In Fig. 4, the relative position of maximum growth $y_d$ is varied from 0.1 (Fig. 4(a)) to 0.9 (Fig. 4(d)). The maximum width migrates from the lower to the upper end of the leaf. These forms change through the sequence deltoid – ovate – obovate – obdeltoid. Note that Fig. 4(a) and Fig. 4(d) are exact inverses due to the symmetry of the growth function, Eq. (3). Once again, only one parameter is being varied continuously.

In Fig. 5, the exponent k is varied. For small k = 1.0 in Fig. 5(a) the leaf margin is concave to the midrib. As k increases to 9.0 in Fig. 5(d), the margin becomes convex and more elongate. The forms proceed through concave, rhombic, orbiculate, and oblong.

In Fig. 6, the slope of the veins is varied from a negative value to a positive value. The forms vary from cordate or reniform through ovate, obovate, and obcordate or winged.

Dissected leaves can be demonstrated with a nonzero value of the $a$ parameter in Eq. (4). For rounded lobes, the exponent $n$ can be set to 2.0, a Gaussian exponential function. In Fig. 7(a), $a = 0$, and there are no lobes. As $a$ increases, the growth rate of the intervein cell rows is reduced, and the lobes increase in size. Strong dissection of the leaf is shown in Fig. 7(d). It is interesting that among the oaks (*Quercus*), this sequence is found, and extensive hybridization among species produces a complete spectrum of leaves from entire (7(a)) to strongly dissected (7(d)).

For sharply pointed lobes, the exponent $n$ is changed to 1.0, a simple exponential with a discontinuity in slope at each vein tip. The equivalent to the Fig. 7 sequence is shown in Fig. 8. While the white oaks have rounded lobes as in Fig. 7, the red oaks have pointed lobes, as in Fig. 8.

More complex leaves such as the palmate leaf (maple (*Acer*), sweet gum (*Liquidambar*), sycamore (*Platanus*)) require a more complex model, since the simple midrib organizing principle is replaced by a self-similar organization in which the secondary veins replicate the midrib with autonomous growth characteristics. However, a rough approximation to maple leaves is shown in Fig. 9. Here the vein structure is set to resemble the splayed lobes in the maple, the exponent $n$ is set to 1.0, and the exponential factor $a$ is varied from 0.03 to 0.

As the *a* value decreases, the growth algorithm "fills in" the spaces between the lobes and the leaf becomes a rounded shape typical of the entire leaves of many other plant species. Fig. 9 approximates the leaf shapes found in various maple species.

## V. DISCUSSION

The conclusion of this work is that a simple growth algorithm can produce leaf shapes resembling those observed in nature. With only a few parameters, the 28 forms produced in Figs. 3 – 9 were generated. This suggests that a more biologically-based classification of leaves would refer to growth rate parameters rather than final shape. The model can easily produce shapes not found in nature. Perusal of many regional floras has not turned up any instance of Fig. 5(a), for example. The model could therefore be useful in exploring the "morphospace" of possible leaf forms [19].

What might be the biophysical significance of the proposed growth algorithm? It seems likely that genetic factors in leaf growth control not final shape but spatially varied growth rates [20]. It is also plausible that the growth rate for each tissue element is determined by a positional information mechanism established at an early stage in the leaf primordium, and is expressed throughout the expansion of the leaf. The final leaf shape is the "accidental" outcome of the growth rate law, but natural selection actively eliminates the most inefficient forms. Exotic leaf shapes produced by plant breeders but not found in natural populations are evidence of this selection process.

Even under natural selection, however, the variability of leaf shapes is large. Two possible reasons are suggested by the above results: (1) changing the point of leaf maturation would produce allometric shape variation, and (2) the sensitivity of leaf shape to small changes in the growth algorithm means that natural variations in this algorithm could induce shape variability. If the proposed positional information mechanism is actually a cascade of gene products whose effects are mediated by chemical kinetic and transport processes, then any external influence on these processes would also create variability in final leaf shape. Heteroblastic leaves found on a single plant are likely to be the result of such influences. Dissected leaves appear to be especially variable. The mulberry (*Morus*) and sassafras (*Sassafras*) may have entire and dissected leaves on the same tree. Perhaps the biophysical process modeled by the *a* parameter in Eq. (4) is mediated by environment-sensitive variables.

The growth-algorithm model was deliberately made very simple. A typical simulation takes less than 1 second of computer time. But it can be elaborated to include a more general non-uniform growth law, left-right growth asymmetry, nonlinear veins, palmate vein patterns, and compound leaves. By relaxing the assumptions made, it should be possible to include more types of leaf shapes and to improve the accuracy of the agreement with observed shapes.

## ACKNOWLEDGMENT

This work was performed during 2006 under the auspices of the U. S. Department of Energy by the University of California, Lawrence Livermore National Laboratory under contract No. W-7405-Eng-48.

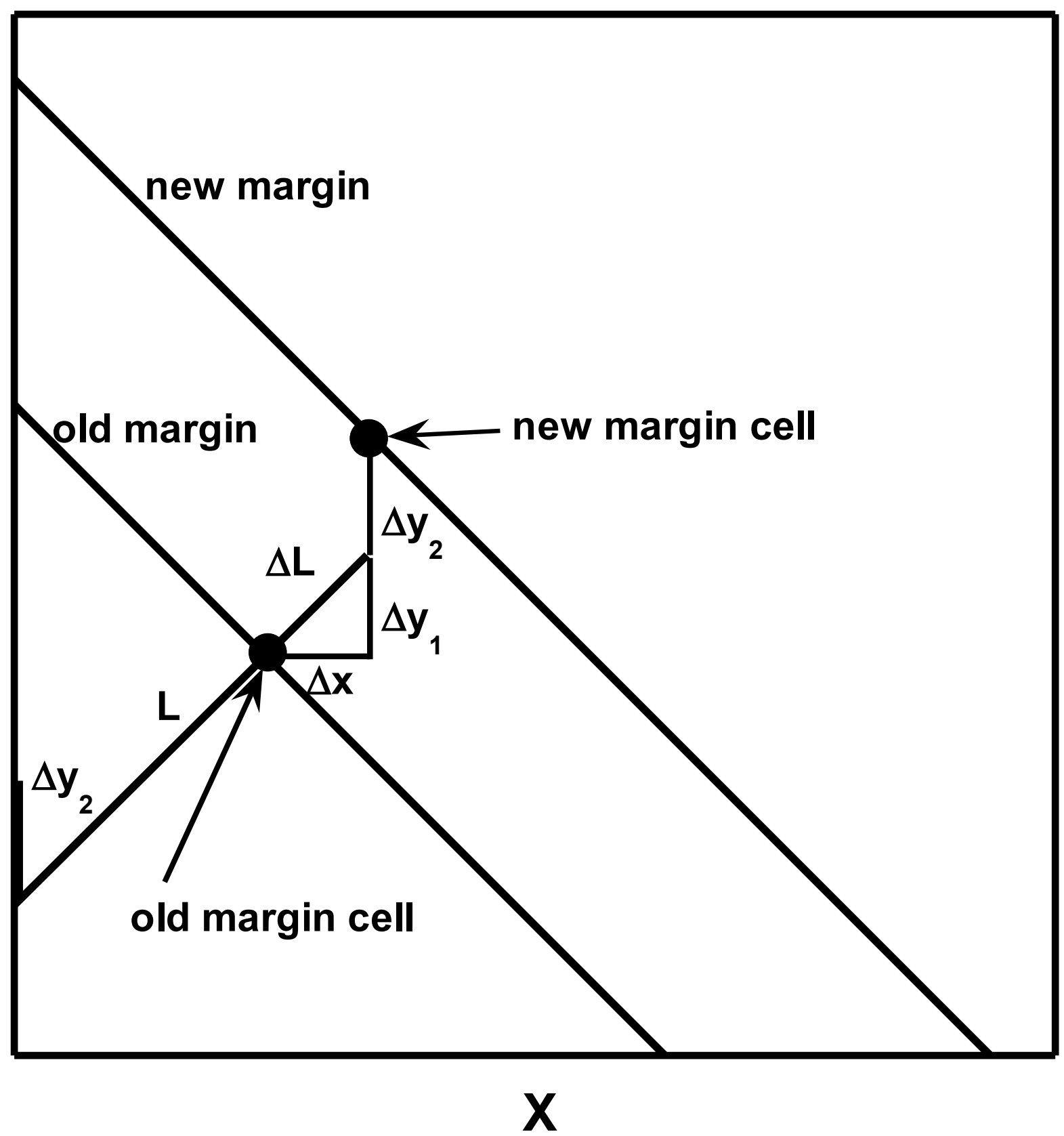

FIG. 1. A schematic outline of the growth algorithm. The new position of a margin cell due to cell row growth is indicated by the combined cell row elongation and y elongation.

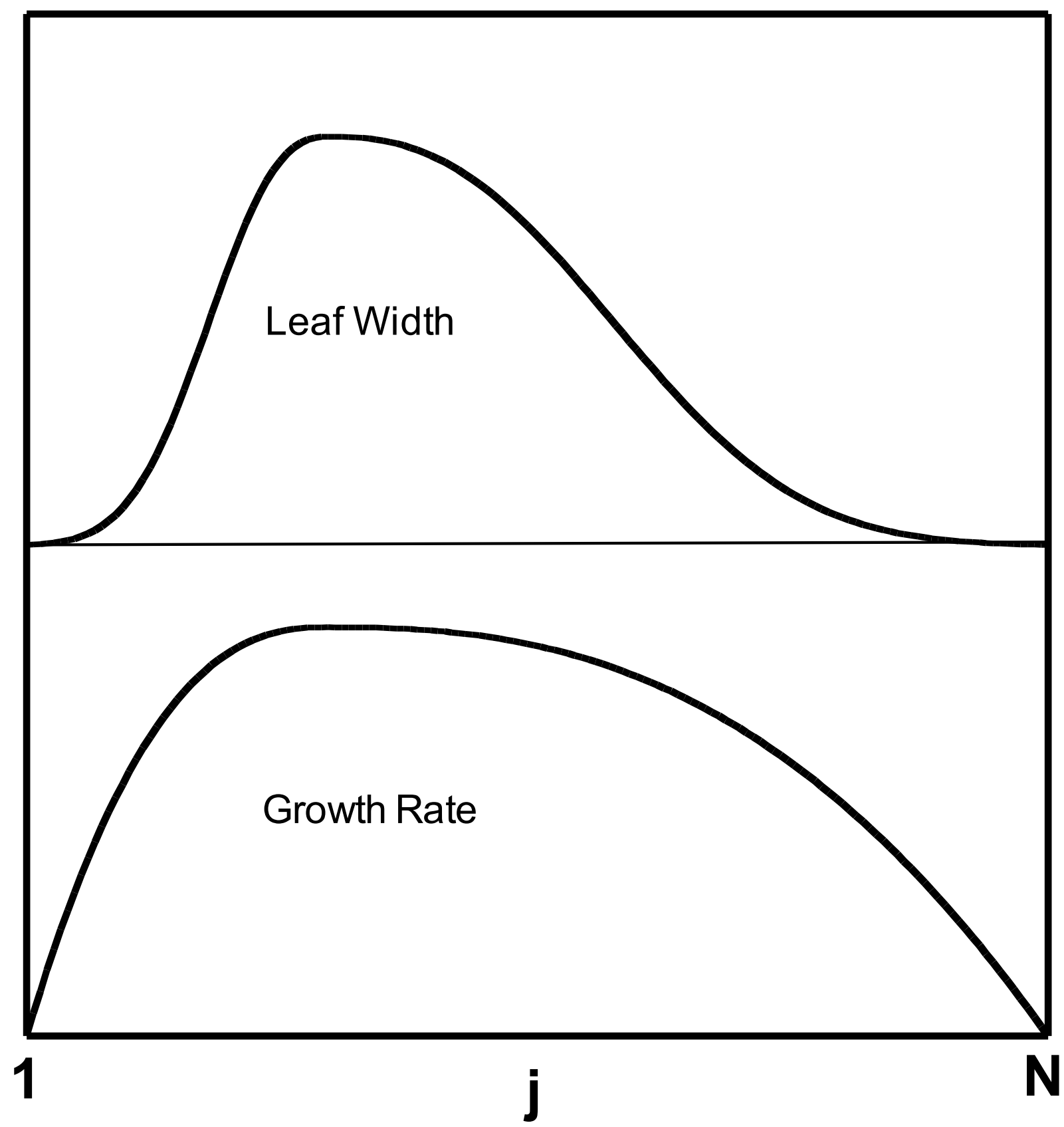

FIG. 2. The growth rate $g_L$ (lower panel) is compared with the mature leaf outline (upper panel) as a function of the index j. The exponential growth amplifies the shape variation in $g_L$. Here $k = 2.5$, $y_d = 0.3$, $R = 1.20$, and $s = 0$.

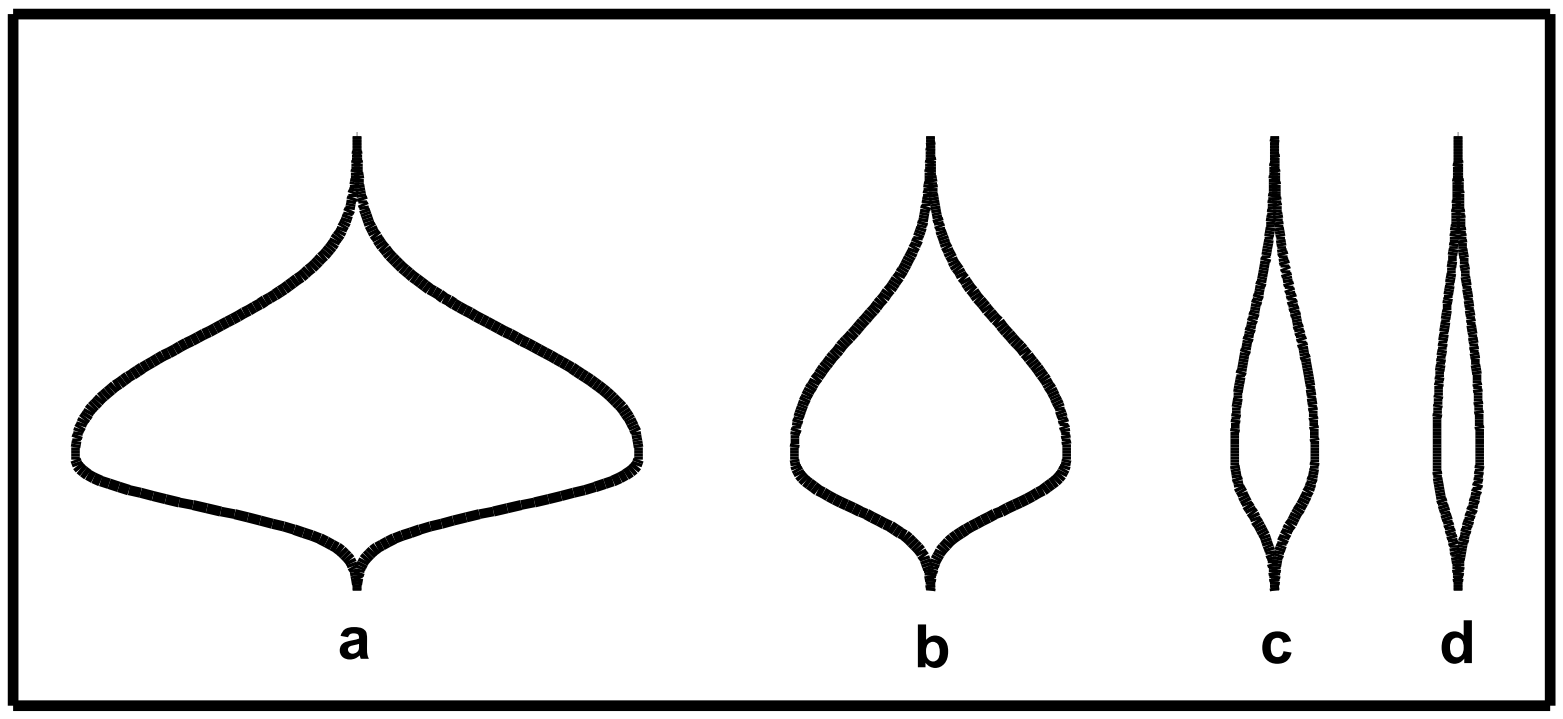

FIG. 3. Four leaf shapes produced by setting $k = 2.5$, $y_d = 0.3$, $s = 0$, and varying the ratio $R = g_y/g_{L0}$ . (a) $R = 1.07$, broadly ovate. (b) $R = 1.20$, ovate. Shapes (a) and (b) are seen in poplars (*Populus*). (c) $R = 1.50$, broadly lanceolate. (d) $R = 1.71$, lanceolate. Shapes (c) and (d) are seen in willows (*Salix*).

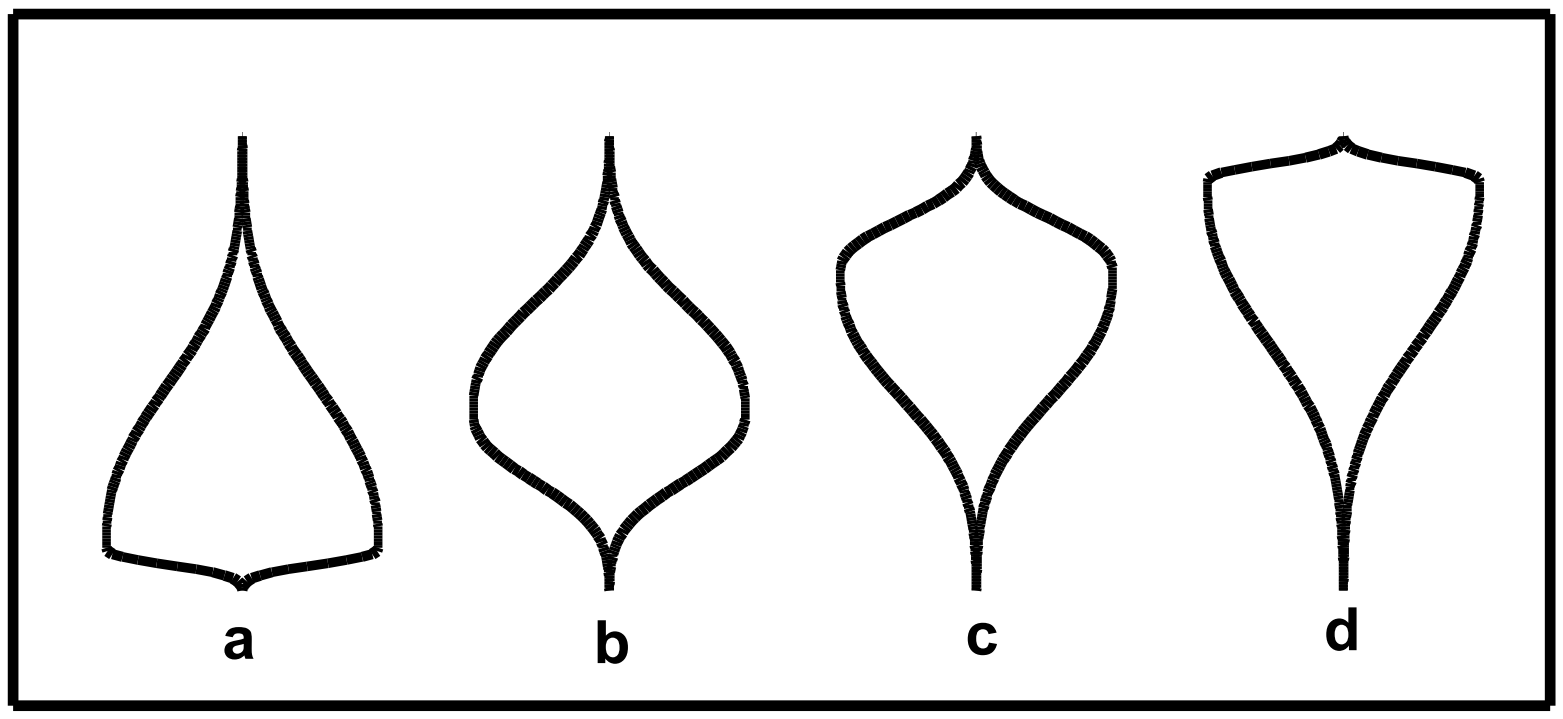

FIG. 4. Four leaf shapes produced by setting R = 1.2, k = 2.5, s = 0, and varying the fractional position of maximum growth $y_d$. (a) $y_d = 0.1$, deltoid, seen in birch (*Betula*) and cocklebur (*Xanthium*). (b) $y_d = 0.4$, ovate, seen in poplar (*Populus*), alder (*Alnus*), and elm (*Ulmus*). (c) $y_d = 0.7$, obovate, seen in magnolia (*Magnolia*). (d) $y_d = 0.9$, obdeltoid, seen in ginkgo (*Ginkgo biloba*).

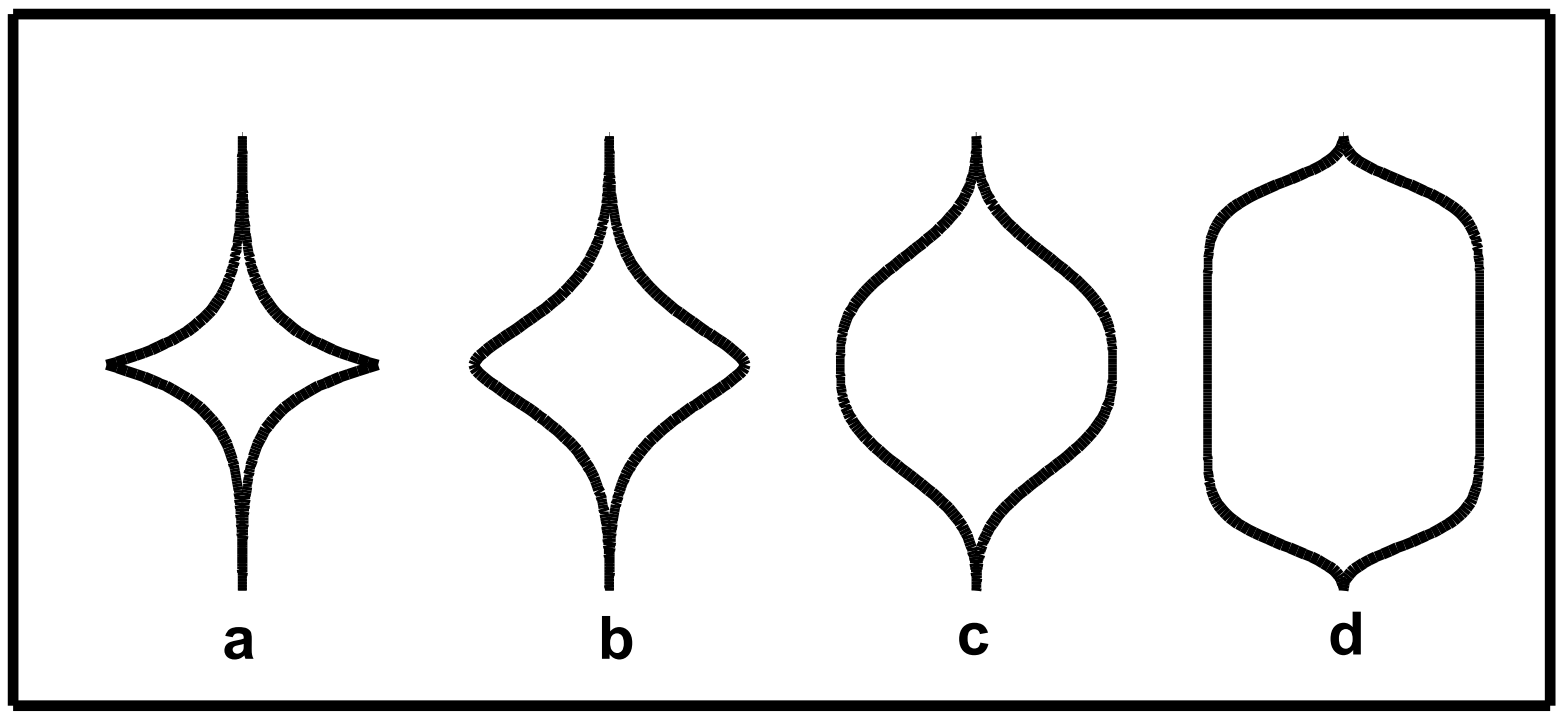

FIG. 5. Four leaf shapes produced by setting R = 1.2, $y_d$ = 0.5, s = 0, and varying the exponent k. (a) k = 1.0, concave. (b) k = 1.5, rhombic, seen in the South American species *Jodina rhombifolia*. (c) k = 3.0, orbiculate, seen in oregano (*Origanum*) and blackberry (*Rubus*). (d) k = 9.0, oblong, seen in buckthorn (*Rhamnus*) and fig (*Ficus*).

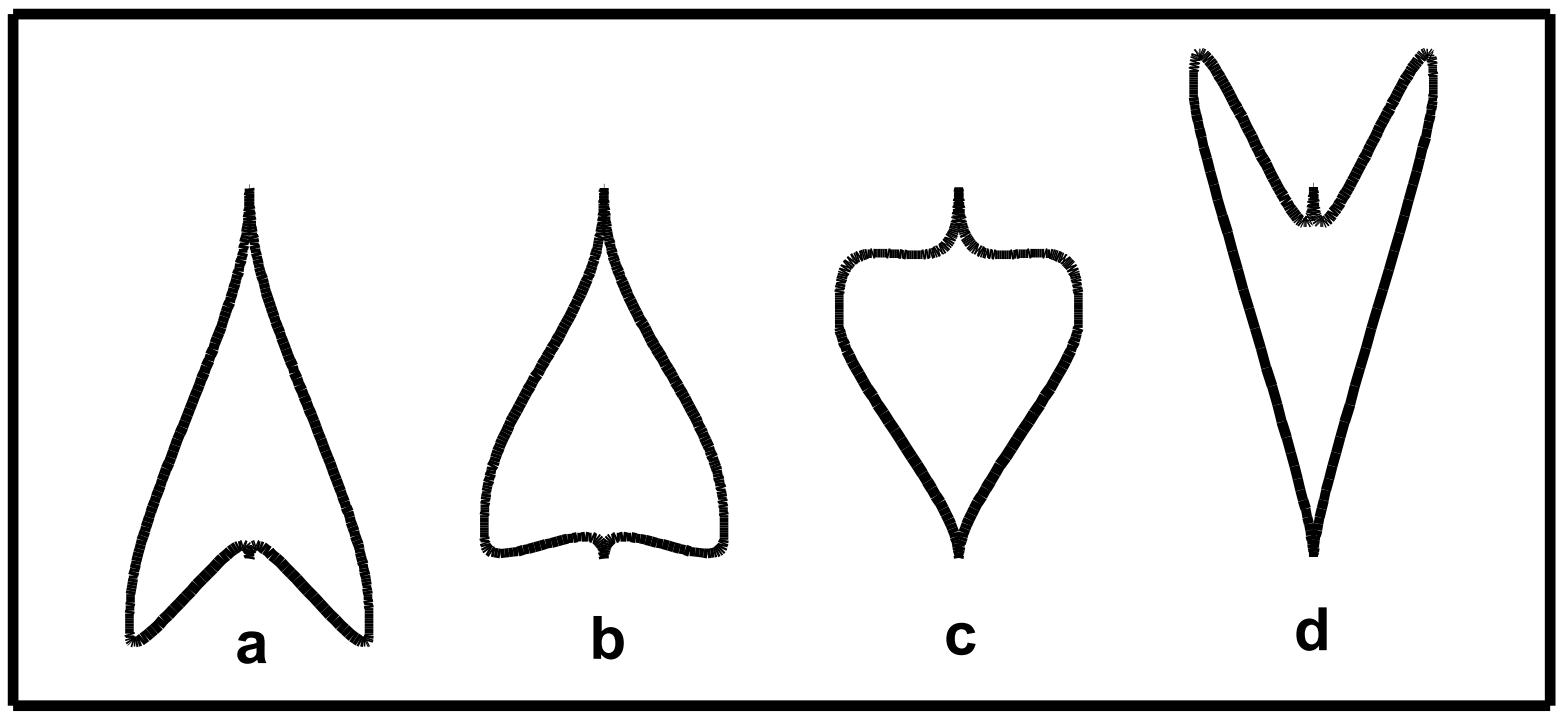

FIG. 6. Four leaf shapes produced by setting R = 1.2, $y_d$ = 0.5, k = 3.0, and varying the slope s of the veins. (a) s = –1.5, cordate or reniform, seen in arrowhead (*Sagittaria*). (b) s = –0.7, deltoid, seen in morning glory (*Calystegia*). (c) s = +1.1, obdeltoid. (d) s = +3.0, winged, seen in bauhinias (*Bauhinia*).

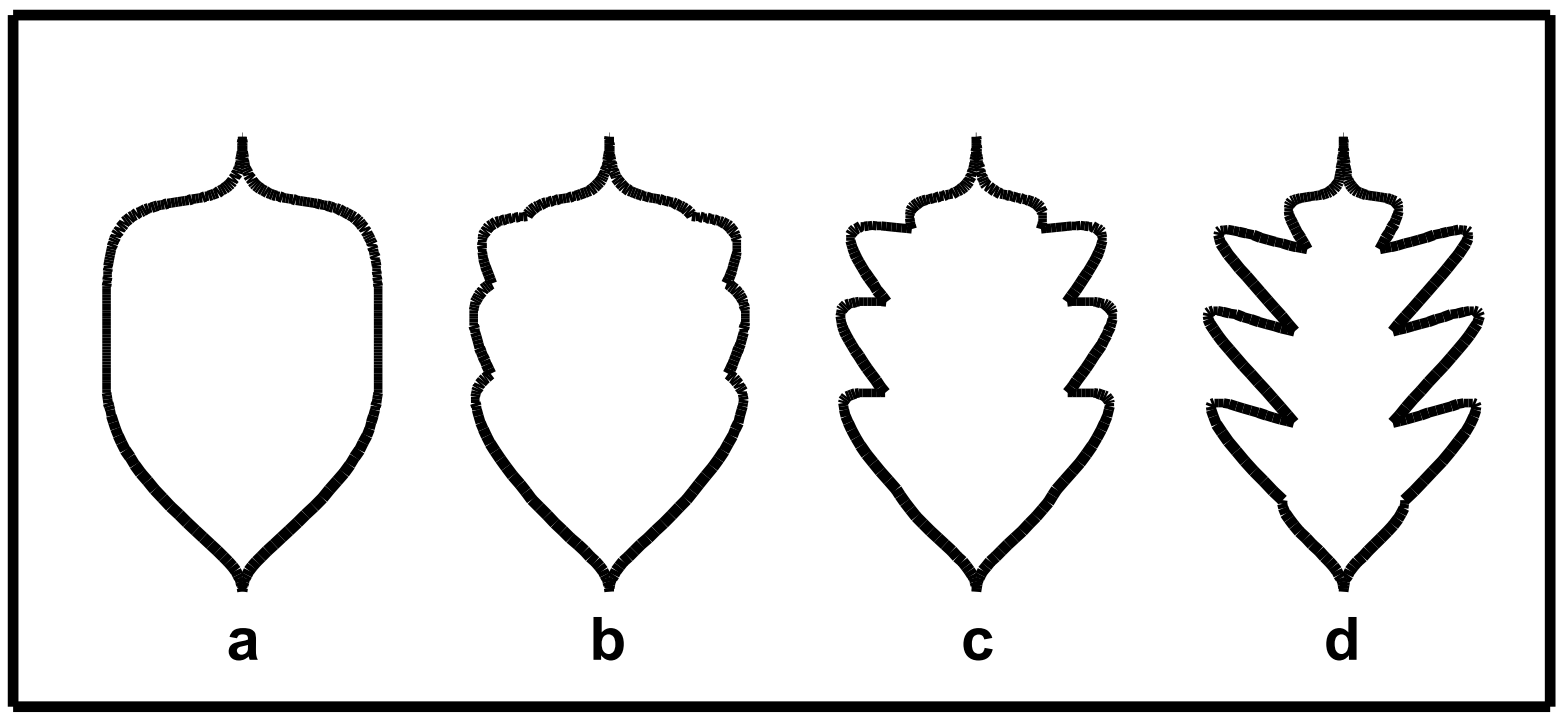

FIG. 7. Four oak (*Quercus*) leaf shapes produced by setting R = 1.2, $y_d$ = 0.3, k = 3.0, s = 0.7, n = 2, and varying the exponent factor *a*. (a) *a* = 0, entire leaf, seen in *Q. myrtifolia*. (b) *a* = 0.0001, serrated leaf, seen in *Q. chapmanii*. (c) *a* = 0.0003, toothed leaf, seen in *Q. alba*. (d) *a* = 0.0008, dissected leaf, seen in *Q. lobata*.

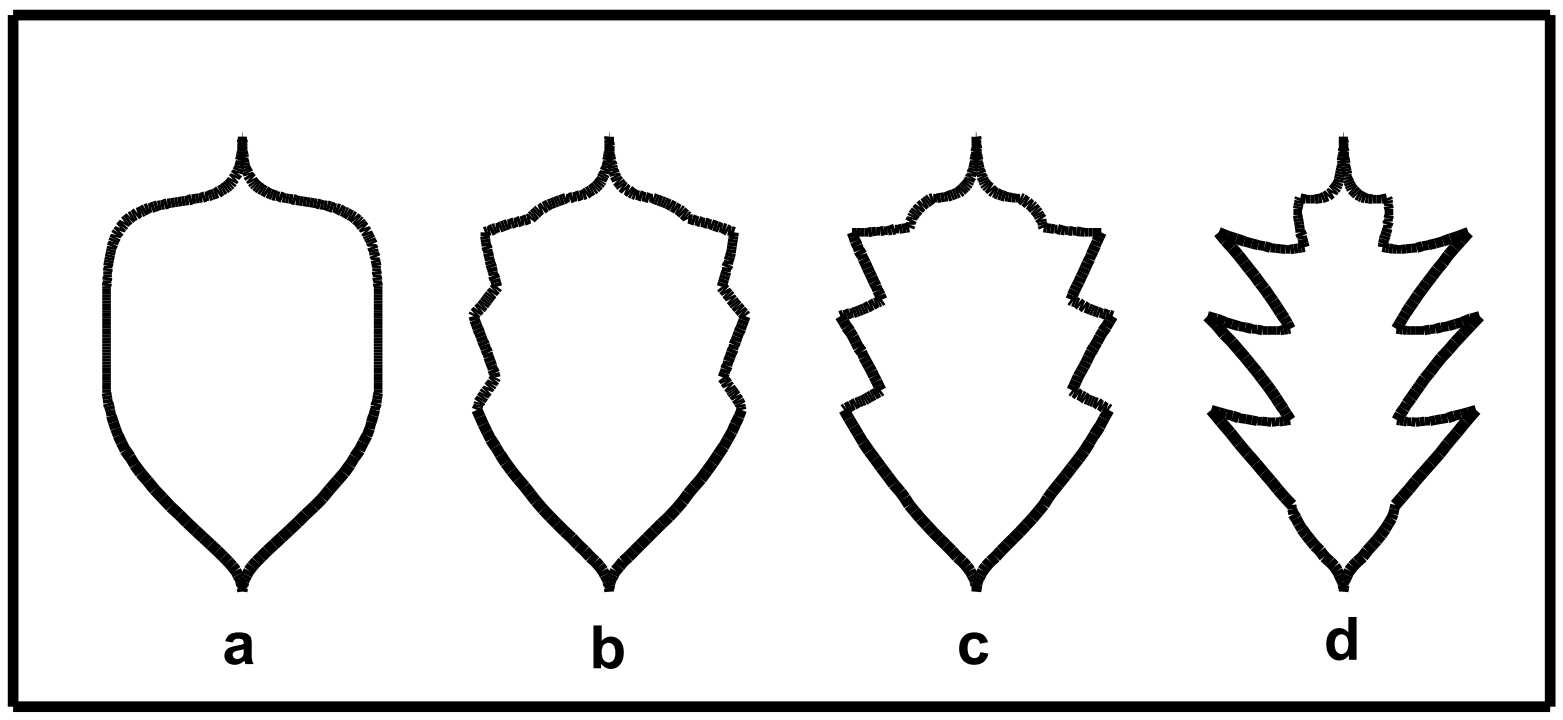

FIG. 8. Four oak (*Quercus*) leaf patterns produced by setting R = 1.2, $y_d$ = 0.3, k = 3.0, s = 0.7, n = 1, and varying the exponent factor *a*. (a) *a* = 0, entire leaf, seen in *Q. myrtifolia*. (b) *a* = 0.002, serrated leaf, seen in *Q. agrifolia*. (c) *a* = 0.004, toothed leaf, seen in *Q. engelmanii*. (d) *a* = 0.011, dissected leaf, seen in *Q. rubra*.

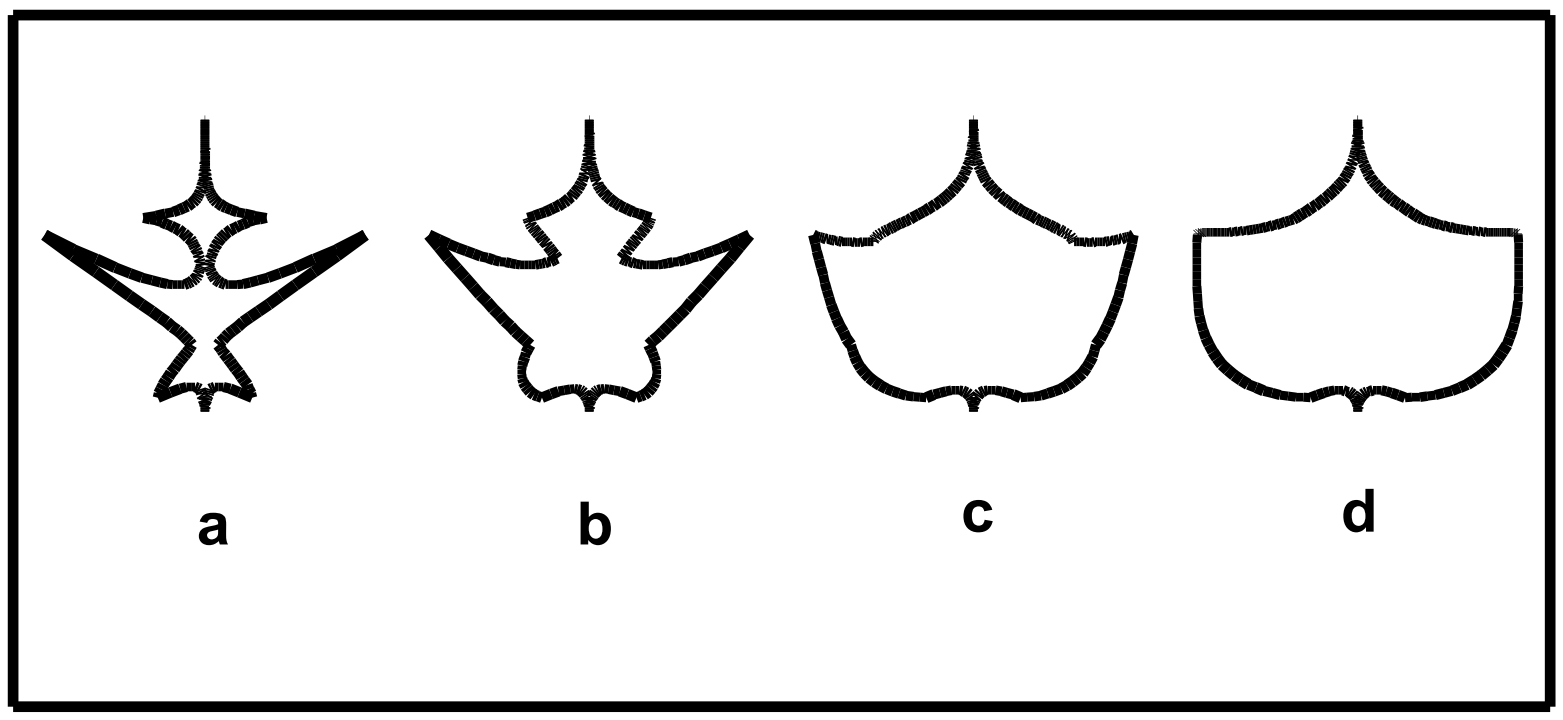

FIG. 9. Four maple (*Acer*) leaf patterns produced by setting R = 1.2, $y_d$ = 0.3, k = 3.0, n = 1 and varying the exponent factor *a*. (a) *a* = 0.030, highly dissected leaf with narrow lobes, seen in *A. elegantum*. (b) *a* = 0.010, weaker dissection and larger lobes, seen in *A. campestre*. (c) *a* = 0.002, nearly continuous outline, seen in *A. conspicuum*. (d) *a* = 0, entire leaf with no lobes, seen in *A. caesium*.